\input{aipcheck}

\documentclass[
    ,final            % use final for the camera ready runs
  ]
  {aipproc}

\layoutstyle{6x9}

\begin{document}

\title{ Possible saturation effects at HERA and LHC\\ in the 
$k_T-$factorization approach}


\author{A.V. Kotikov}{
  address={N.N. Bogoliubov Laboratory of Theor. Physics,
Joint Institute for Nuclear Research,\\
 141980 Dubna, Russia}
}

\author{A.V. Lipatov}{
  address={D.V. Skobeltsyn Institute of Nuclear Physics,
M.V. Lomonosov Moscow State University,\\
 119992 Moscow, Russia}
}

\author{\underline{N.P. Zotov}}{
 address={D.V. Skobeltsyn Institute of Nuclear Physics,
M.V. Lomonosov Moscow State University,\\
 119992 Moscow, Russia}
%  ,altaddress={<author1 address>} % additional visiting address
}

\begin{abstract}
  We consider possible saturation effects in the structure function
$F_L$ at fixed W in HERA energy range and in $b\bar b-$production
at LHC energies in the framework of $k_T-$facrorization approach.
\end{abstract}

\maketitle

%%%%%%%%%%%%%%%%%%%%%%%%%%%%%%%%%%%%%%%%%%%%
%% MAINMATTER
%%%%%%%%%%%%%%%%%%%%%%%%%%%%%%%%%%%%%%%%%%%%

\section{Introduction}

  Remarkable progress in recent years has resulted from the observation
that the gluon density in a proton at small $x$ grows as $x$ decreases.
Both DGLAP and BFKL evolution equations predict this
rapid growth of
the parton densities, thus demostrating the triumph of perturbative QCD.
It is, however, clear that this growth cannot continue for ever, because
it would violate the unitarity constraint~\cite{GLR}. Consequently, the
parton evolution dynamics must change at some point, and a new
phenomenon must come into play.
 As the gluon density increases, non-linear parton interactons are
expected to become more and more important, resulting eventually in the
slowdown of
the parton density growth (known under the name of "saturation effect")
~\cite{GLR,MQ}. The underlying physics can be described by the
non-linear Balitsky-Kovchgov (BK) equation~\cite{BK}. It is expected
that these nonlinear interactions
lead to an equilibrium-like system of partons with some definite value
of the average transverse momentum $k_T$ and the corresponding
saturation scale $Q_s(x)$. This equilibrium-like system is the so called
Color Glass Condensate (CGC)~\cite{GMcL}.
Since the saturation scale increases with
decreasing of $x$, one may expect that the  saturation effects will be
more clear at the LHC energies.  In the preasymtotic region ($k_T \geq
Q_s(x)$) of energies
the heavy quark production is described more adequately~\cite{GV} by
the so called $k_T-$factorization approach~\cite{GLR,LRSS}.
At $k_T < Q_s(x)$ the $k_T-$factorization approach gives us a chance
to account for saturation effects (if they are under control the BK
equation) using scale properties of the dipole model, which is
equivalent the $k_T$ factorization~\cite{NN}.

Here we demonstrate a description of possible saturation effects  in
the framework of $k_T-$factorization approach. As example of that we
consider the structure function $F_L$ at fixed W and low $Q^2$ in
 HERA energy range and $b\bar b-$quark production at LHC energies.

\section{Theoretical framework and numerical results}
 
\subsection{The SF $F_L$ at fixed W in HERA range}

\begin{figure}
%\special{psfile=FLW-ll.eps
%voffset=50 vscale=40
%hscale=50 hoffset=-200 angle=-90}
\includegraphics[width=0.7\linewidth]{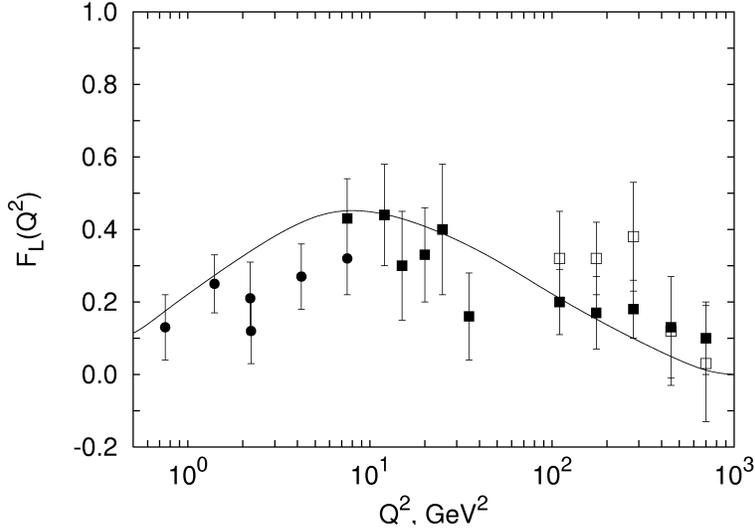}
  \caption{The $Q^2$ dependence of $F_L(x, Q^2)$ (at fixed $W$ = 276
GeV). The experimental points are from \cite{H1}.
Solid curve is the result of the $k_T-$factorization approach
 with the GLLM unintegrated gluon distribution from~\cite{LL}.
\label{fig:flw}}
\end{figure}

 The SF $F_{L}(x,Q^2)$ is driven at small $x$ primarily
by gluons and in the $k_T$-factorization approach
it is related in the following 
way to the unintegrated distribution
$\Phi_g(x,k^2_{\bot})$:
\begin{eqnarray}
F_{L}(x,Q^2) ~=~\int^1_{x} \frac{dz}{z} \int^{Q^2}
dk^2_{\bot} \sum_{i=u,d,s,c} e^2_i
\cdot \hat C^g_{L}(x/z,Q^2,m_i^2,k^2_{\bot})~ \Phi_g(z, k^2_{\bot}),
 \label{d1}
\end{eqnarray}
where $e^2_i$ are charge squares of active quarks.
 The function $\hat C^g_{L}(x,Q^2,m_i^2,k^2_{\bot})$
can be regarded as  SF of the
off-shell gluons with virtuality $k^2_{\bot}$ (hereafter we call them
{\it hard structure function } by analogy with similar
relations between cross-section and hard
cross-section). They are described by the sum of the quark
box (and crossed box) diagram contribution to the
photon-gluon interaction (see, for example, Fig. 1 in \cite{KLZ1}).
To calculate the longitudinal SF $F_{L}(x,Q^2)$ we used the hard SF 
$\hat C^g_{L}(x,Q^2,m_i^2,k^2_{\bot})$ obtained in Ref.~\cite{KLZ1}
and different unintegrated gluon distributions. \footnote{Here we are 
interested 
in the parametrizations of unintegrated gluon functions which take into 
account sturations effects.}
Notice that the $k^2_{\bot}$-integral in Eq. (\ref{d1})
can be divergent at lower limit,
at least for some parameterizations of $\Phi_g(x,k^2_{\bot})$.
To overcome the problem we change the low $Q^2$ asymptotics of
the QCD coupling constant within hard structure functions.
We applied the so called "soft" version of "freezing" procedure 
\cite{NikoZa},which contains the
shift $Q^2 \to Q^2 + M^2$, where $M$ is an additional
scale, which strongly modifies the infrared $\alpha_s$ properties.
For massless produced quarks, $\rho$-meson mass $m_{\rho}$ is usually
taken as the $M$ value, i.e. $M=m_{\rho}$.
In the case of massive quarks with  mass $m_i$, the $M=2m_{i}$ value
is usually used. We calculate the SF $F_L$ as the sum of two types of 
contributions -
the charm quark one $F^c_L$ and the light quark one $F^{l}_L$:
\begin{eqnarray}
F_L  ~=~ F^{l}_L + F^{c}_L.
\end{eqnarray}
For the $F^{l}_L$ part we used the massless limit of hard SF and 
resticted ourselves to the modification of the argument in the strong
coupling constant of the hard SF only.
We have shown~\cite{KLZ2} that our $k_T$-factorization
 results are in good agreement with the data for large and small
part of the $Q^2$ range. However, there was some disagreement
 between the data and theoretical predictions at $Q^2 \sim 3$ GeV$^2$.
The disagreement  comes from two possible reasons:  additional 
higher-twist
contributions, which are important at low $Q^2$ values\footnote{Some 
part of higher-twist contributions was took into account by the
"freezing" procedure.}, or/and  NLO QCD corrections.
\begin{figure}
\includegraphics[width=0.5\linewidth]{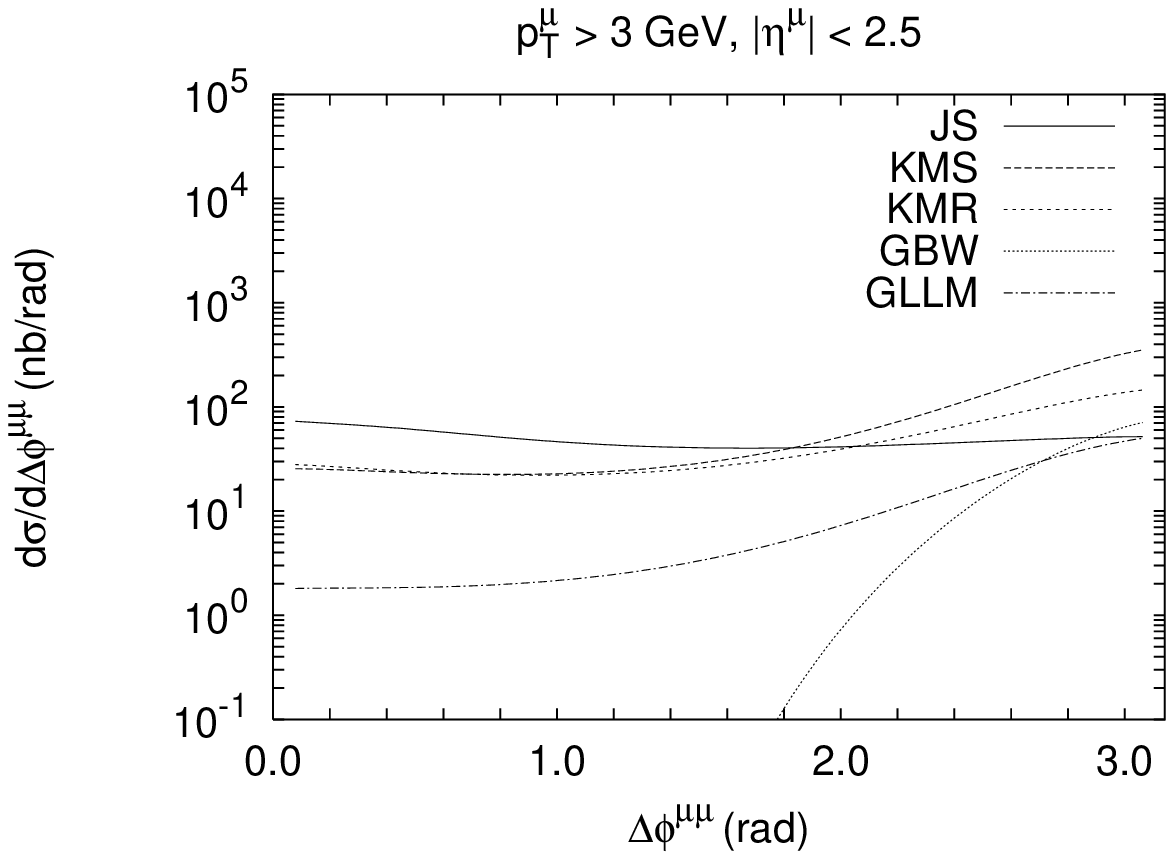}
\includegraphics[width=0.5\linewidth]{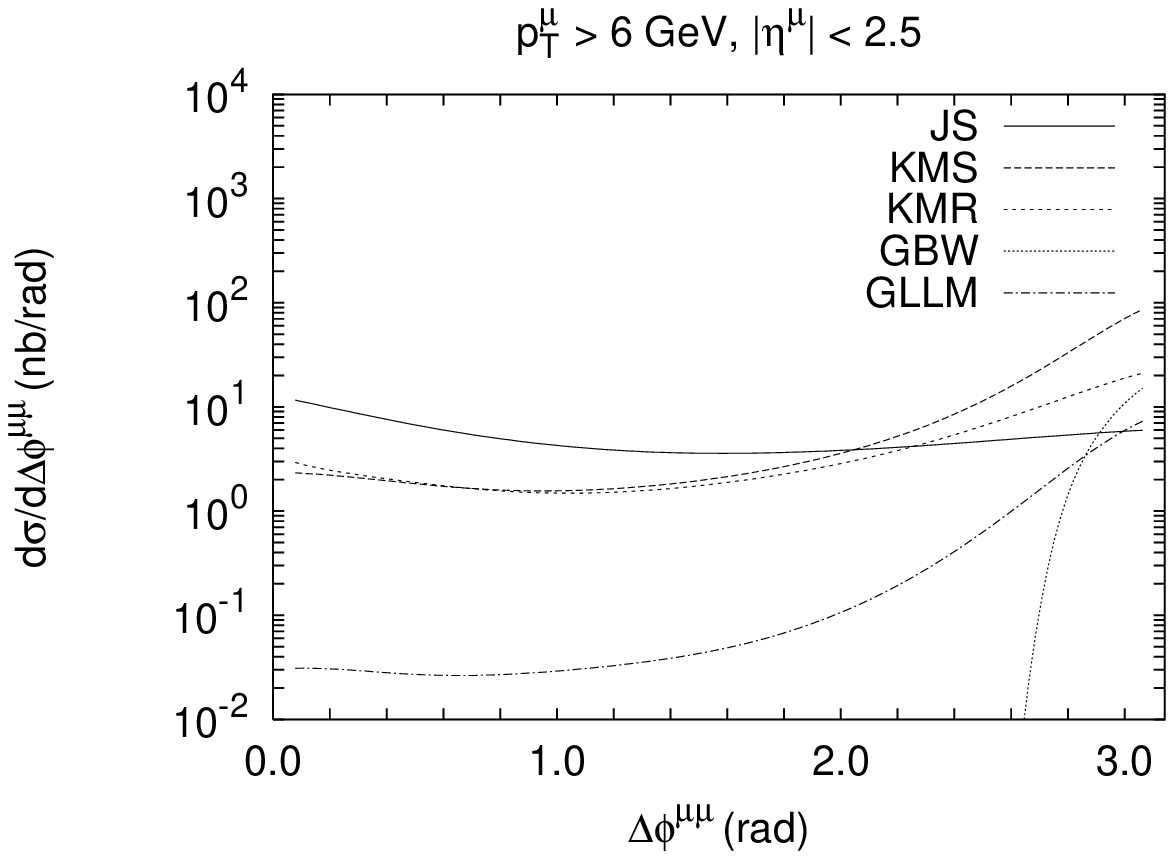}
\vspace*{0.5mm}
\caption{Azimutal muon-muon correlations at LHC with different 
unintegrated gluon distributions
\label{fig:lhc1}}
\end{figure}
 
   It was shown that the saturation (non-linear QCD) approaches contain
information of all orders in $1/Q^2$, they resum higher-twist
contributions~\cite{Bar}. The analysis of the behaviour of the
longitudinal structure function $F_L(x, Q^2)$ in the saturatation models
was done in Ref.~\cite{Mach}\footnote{N.Z. thanks M.V.T. Machado for
useful discussion of this problem.}. In Fig.1 we demonstrate 
our
$k_T-$factorization description of $F_L(Q^2)$ at fixed $W$ with the
unintegrated gluon distribution proposed in Ref.~\cite{LL} which
takes into account non-linear (saturation) effects. We see very well 
description of H1 experimental data at fixed W in all $Q^2$ region.

\subsection{Saturation effects in $b\bar b-$production at LHC}
 
 It was shown that the data on $b\bar b$ azimutal correlations at 
Tevatron (measured as the decay muon ones, 
$d\sigma/d\Delta\phi^{\mu\mu}$) are much more informative to  
distinguish different unintegrated gluon distributions~\cite{BZL}.
In Fig. 2 we show our predictions for the $b\bar b$ azimutal 
correlations at LHC obtained with different  unintegrated gluon 
distributions (see, for example,~\cite{Smallx}). The GBW and GLLM
ones take into account saturation effects. We see that  the 
latter two unintegrated gluon densities reduce to different
behaviour of differential cross section 
$d\sigma/d\Delta\phi^{\mu\mu}$. Fig. 3 displays the total transverse 
momentum distribution, $d\sigma/dp_T^{b\bar b}$, where $p_T^{b\bar b} =
p_T^b + p_T^{\bar b}$, at LHC energy obtained with the same  
unintegrated gluon densities. In this case the difference between 
the results with usual unintegrated gluon densities and ones
obtained with account of the  non-linear (saturation) effects is more 
dramatic.

\begin{figure}
\includegraphics[width=0.7\linewidth]{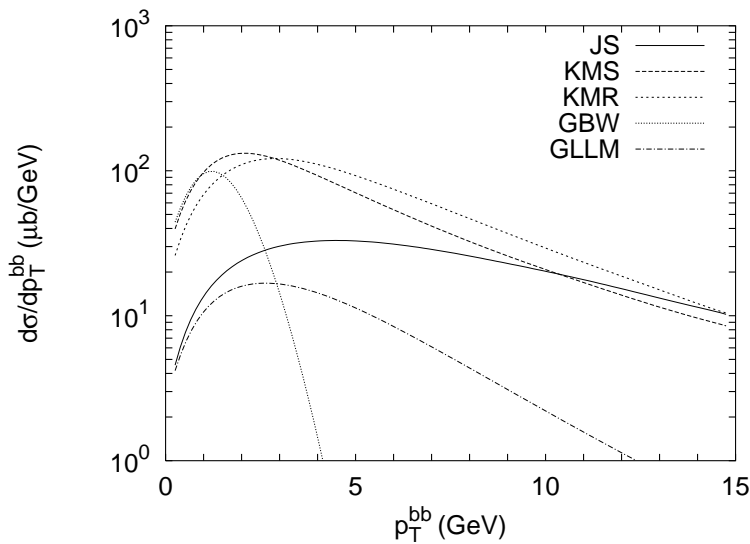}
\vspace*{0.5mm}
\caption{The total transverse momentun distribution at LHC
with different unintegrated gluon distributions
\label{fig:lhc2}}
\end{figure}

\section{Conclusions}

We  considered in the framework of $k_T-$factorization approach
possible manifestation of the saturation effects at HERA and LHC.
We shown that account of these effects (in particle, by the GLLM 
unintegrated gluon distribution) improves to a marked degree description 
of the $F_L$ data at low $Q^2$ at HERA energy. We demonstrated also that
the experimental data for $b\bar b$ azimutal correlations and
$p_T^{b\bar b}-$distribution at LHC will give us additional 
possibilities to study non-linear (saturation) effects.
 
\begin{theacknowledgments}
 N.Z. thanks DESY directorate for financial support and
Organizing Committee for
 hostly and friendly atmosphere during the Symposium.
\end{theacknowledgments}
%\end{document}


%\bibliography{ismdZot}

%\IfFileExists{\ismdZot.bbl}{}
%  \typeout{******************************************}
%  \typeout{** Please run "bibtex \ismdZot" to optain}
%  \typeout{** the bibliography and then re-run LaTeX}
%  \typeout{** twice to fix the references!}
%  \typeout{******************************************}

%\endinput
\end{document}